\documentclass[twocolumn,showkeys,show pacs,preprintnumbers,floatfix,amsmath,a4paper,amssymb,nofootinbib]{revtex4}

\usepackage[latin1]{inputenc}
\usepackage{graphicx}
\usepackage{graphicx}
\usepackage{epstopdf}
\usepackage{epsfig}
\usepackage{color}

\usepackage{bm}
\usepackage{graphicx}

\begin{document}

\title{Understanding Zipf's law of word frequencies through sample-space collapse in sentence formation}
\author{Stefan Thurner$^{1,2,3}$, Rudolf Hanel$^1$, Bo Liu$^1$, and Bernat Corominas-Murtra$^1$}

\thanks{stefan.thurner@meduniwien.ac.at}

\affiliation{
$^1$ Section for Science of Complex Systems, CeMSIIS, Medical University of Vienna,  Spitalgasse 23, A-1090, Austria.\\
$^2$ Santa Fe Institute, 1399 Hyde Park Road, Santa Fe, NM 87501, USA. \\
$^3$IIASA, Schlossplatz 1, A-2361 Laxenburg, Austria.}

\begin{abstract}
The formation of sentences is a highly structured and history-dependent process. The probability 
of using a specific word in a sentence strongly depends on the 'history' of word-usage earlier in that sentence.  
We study a simple history-dependent model of text generation assuming that the sample-space of word usage 
reduces along sentence formation, on average. We first show that the model explains the approximate Zipf law found in word frequencies 
as a direct consequence of sample-space reduction. We then empirically quantify the amount of sample-space 
reduction in the sentences of ten famous English books, by analysis of corresponding word-transition tables that capture 
which words can follow any given word in a text. We find a highly nested structure in these transition tables 
and show that this `nestedness' is tightly related to the power law exponents of the observed word frequency distributions. 
With the proposed model it is possible to understand that the nestedness of a text can be the origin of 
the actual scaling exponent, and that deviations from the exact Zipf law can be understood by variations of the degree of 
nestedness on a book-by-book basis. 
On a theoretical level we are able to show that in case of weak nesting, Zipf's law breaks down in a fast transition. 
Unlike previous attempts to understand Zipf's law in language the sample-space reducing model is not based on 
assumptions of multiplicative, preferential, or self-organised critical mechanisms behind language formation, but simply 
used the empirically quantifiable parameter 'nestedness' to understand the statistics of word frequencies. 
\end{abstract}



\maketitle

\section{Introduction}
Written texts show the remarkable feature that the rank ordered distribution of word frequencies follows 
an approximate power law
\begin{equation}
	f(r) \sim r^{-\alpha} \quad, 
\end{equation} 
where $r$ is the rank that is assigned to every word in the text. For most texts, regardless of language, 
time of creation, genre of literature, its purpose, etc., one finds that $\alpha \sim 1$, which 
is referred to as Zipf's law \cite{zipf72}. 
In Fig. 1 the word frequency  is shown for Darwin's text, {\em The origin of species}. 
The quest for an understanding of the origin of this statistical regularity is going on for almost a century. 
Zipf himself offered a qualitative explanation based on the efforts invested in communication events 
by a sender and a receiver \cite{zipf72}. These ideas were later formalised within an information-theoretic framework 
\cite{mandelbrot1953,Harremoes:2001, sole2003, Corominas-Murtra:2011}. 
\begin{figure}[t]
\includegraphics[width= 8.5 cm]{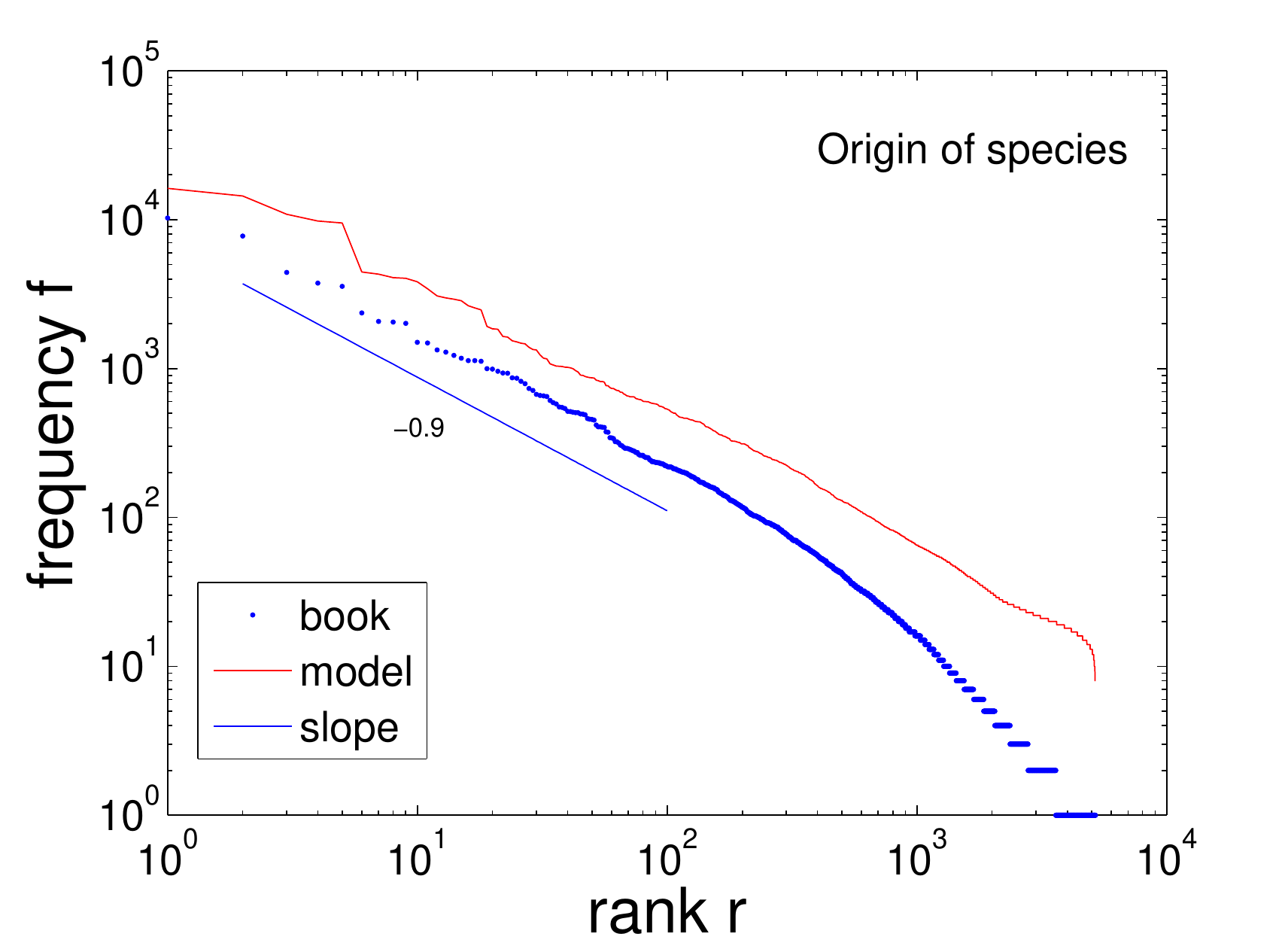}
\caption{Rank ordered distribution of word frequencies for {\em The origin of species} (blue) shows an approximate power law with a slope of 
approximately $\alpha \sim0.9$. The model result (red line) explains not only the power law exponent, 
but also captures details of the distribution. The exponential cutoff can be explained by the randomised version 
of the model.}
\label{fig1}
\end{figure}

\begin{figure*}[t]
\includegraphics[width= 18 cm]{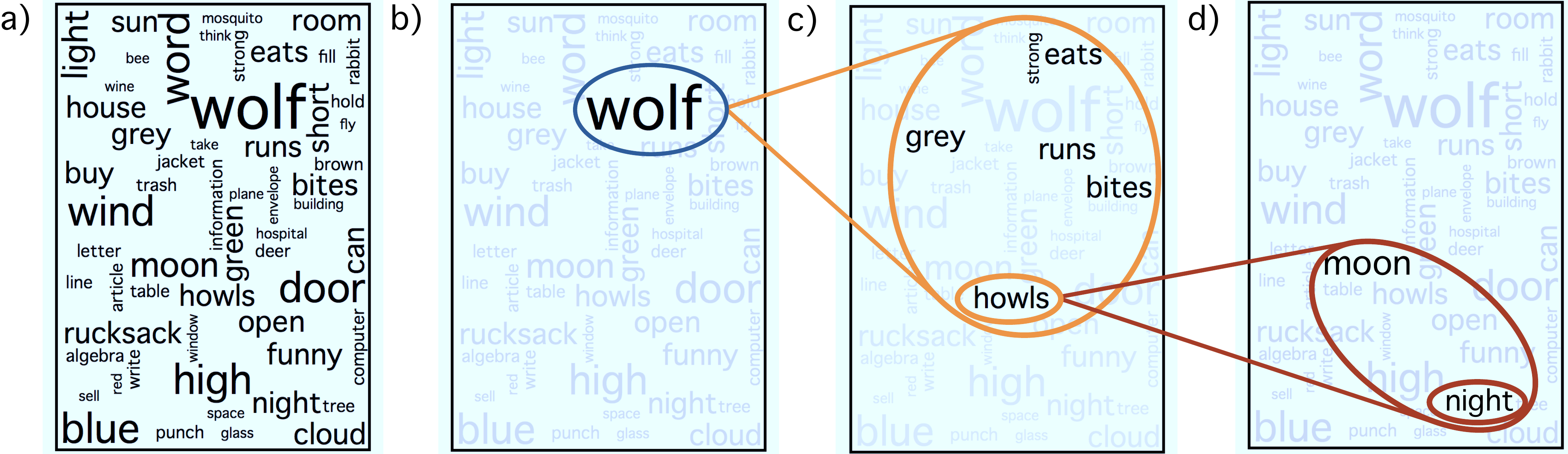}
\caption{Schematic view of nestedness in sentence formation. (a) Among all the potential $N$ words defining the initial sample-space 
we choose ``wolf" (b). This choice restricts the sample-space for the next word  (orange circle) that has to be grammatically and semantically compatible with ``wolf". 
(c) From this set we chose ``howls", which reduces the sample-space again (red circle) (d), since the next word must now be consistent both semantically and 
grammatically with ``The wolf howls". The sequence of words show a nested structure. The effect of sample space collapse is also present 
in the wider context of discourse formation, since a topic and its rhetoric development impose a successive nested constraints of sample space.}
\label{fig:wordSpace}
\end{figure*}

The first quantitative model based on linguistic assumptions about text generation has been proposed by H. Simon \cite{simon:1955}.  
The model assumes that as context emerges in the generation of a text, words that have already appeared in the text are favoured over others. 
By the simple assumption that words that have previously appeared are added to the text with a probability proportional to their previous appearance 
(preferential attachment), and assuming that words that have so far not appeared are added at a constant rate, it is possible to derive Zipf's law, 
given the latter rate is low. This preferential attachment model has been refined by implementing the empirical fact that the rate of appearance 
of new words decreases as the length of texts increases \cite{zanette2005}.  
It has been shown in classical works that random typewriting models can lead to Zipf-like distributions of word frequencies \cite{li1992,miller1957,miller1963}. 
However these works are based on unrealistic assumptions on word-length distributions and lead to unstructured and uninterpretable
texts. 
However, as we will show, syntactic structure, jointly with discourse generation mechanisms, may play an essential a role in the origin of Zipf's law in a realistic context. 
It is important to stress that the detailed statistical study of language properties does not end here; important work beyond Zipf's law has been put forward, see e.g.  
\cite{Kosmidis:2006, Wichmann:2005, serrano2009,zanette2011}. 
Recent studies deal with the detailed dependence of the scaling exponents on the length of the body of text under study
\cite{corral:2013,hisenemies}. 

Zipf's law is not limited to word frequencies but appears in countless, seemingly unrelated, systems and processes \cite{Kawamura:2002}. 
Just to mention a few, it has been found in the statistics of firm sizes \cite{axtell}, city sizes 
\cite{zipf72,  simon:1955,  Makse:1995,Krugman:1996,  Blank:2000,Decker:2007}, the genome \cite{stanley1999},  
family names \cite{Zanette:2001}, income \cite{Pareto:1896, Okuyama:1999}, financial markets \cite{Gabaix:2003}, 
internet file sizes \cite{Reed:2002}, or human behaviour \cite{thurner2012}, for more examples see \cite{Newman:2005}.
There has been tremendous efforts to understand the origin of Zipf's law, and more generally the origin of scaling in complex systems. 
There are three main routes to scaling:  multiplicative processes \cite{simon:1955,mandelbrot1953,Levy:1996}, 
preferential processes \cite{Biham:1999,solar,barabasi}, and self-organised criticality \cite{Bak:1987}.  
Several other mechanisms that are more or less related to these basic routes to scaling have been proposed e.g. in  
\cite{saichev2008,pietronero,thurnertsallis,Corominas-Murtra:2010,Corominas-Murtra:2011,Montroll:1982}. 

Recently a fourth, independent route to scaling has been introduced on the basis of stochastic processes 
that reduce their potential outcomes (sample-space) over time \cite{Zipf1C-MHT}. These are history-dependent random processes that have been studied 
in different contexts in the mathematical literature \cite{kac,clifford}, and more recently in the context of scaling laws \cite{hanel2014,hanel2013}. 
An example of sample-space reducing processes is the following. Think of a set of $N$ dice where 
die number 1 has 1 face, die number 2 has two faces (coin), die number 3 has three faces, and so on. Die number $N$ has $N$ faces. 
Start by picking one of the $N$ dice  at random, say dice number $i$.  Throw it and record the obtained face value, which was say $k$.
Then take die number $k-1$ throw it, get $j$, record $j$, take die number $j-1$, throw it, etc. Keep throwing dice in this way until you throw 1
for the first time. Since there is no die with less than 1 faces, the process ends here. 
 The sequence of recorded face values in the above prescription $(i,k,j,\cdots,1)$, is obviously strictly ordered or nested, $i>k>j>\cdots > 1$. 
In \cite{Zipf1C-MHT}  it was shown rigorously that if this process is repeated many times, the distribution of outcomes (face values $1,2, \cdots, N$)
is an exact Zipf law, i.e. the probability to observe a face value $m$ in the above process (sequence of throws) is  exactly
$P_N(m) =  m^{-1} $, given we start with $N$ dice. 
Note that it is necessary to keep $N$ fixed during the repetitions of the process to obtain the exact Zipf law. 
If $N$ varies during the repetitions, clearly Zipf scaling is present asymptotically for high ranks, however due to the mixing of different $N$, 
deviations from the exact Zipf law will appear for low ranks.

More formally, every die $N$ has a sample-space, denoted by $\Omega_N=\{1,2,\cdots,N\}$, which is the number of potential outcomes, i.e. 
the number of faces of dice $N$.
Throwing these dice in the above way gives rise to a sequence of nested sample-spaces
\begin{equation}
	\Omega_1\subset \Omega_2\subset . . . \subset\Omega_{N} \quad.
\label{eq:Nested}
\end{equation}
The nestedness of sample-spaces in a history-dependent sequence is at the heart of the origin of scaling laws in this type of processes.  
For details see \cite{Zipf1C-MHT} where it is also shown that if noise is added to the history-dependent processes, 
the scaling law, $P_N(m) \propto m^{-\lambda}$  is obtained, where $0<1-\lambda < 1$, is the noise level. 

In this paper we present a derivation of Zipf's law of word frequencies, based on a simple model for sentence/discourse formation. 
The model is motivated by the observation that the process of forming a sentence -- or more generally a discourse -- 
is a history-dependent sample-space reducing process. 
Words are not randomly drawn from the sample-space of all possible words, but are used in strict relations to each other. 
The usage of specific words in a sentence highly restricts the usage for consecutive words, leading to a nesting (or sample-space reducing) 
process, similar to the one described above.    
Sample-space collapse in texts is necessary to convey meaningful information. Otherwise, any interpretation, 
even in metaphoric or poetic terms, would become impossible. 
Let us make the point more concrete with an example for the formation of a sentence, where both grammatical and 
contextual constraints (that reduce sample-space) are at work, Fig. 
2. We form the sentence: 
``The wolf howls in the night''. In principle the first word ``The wolf" (ignoring articles and prepositions for the moment) can be drawn from 
all possible words. Assume there exist $N$ possible words, and denote the respective sample-space by $\Omega_N=\{1,2,\cdots,N\}$, 
where each number now stands for one word.  
This is schematically illustrated in Fig. 
2(a). Given that we chose ``The wolf" from $\Omega_N=\{1,2,\cdots,N\}$, 
Fig. 
2(b), the next word will now (usually) not be chosen from $\Omega_N=\{1,2,\cdots,N\}$, but from a subset of it, 
Fig. 
2(c). Imagine that the subset contains $L$ words, we have $\Omega_L \subset \Omega_N$. Typically we expect 
the subset to contain words that are associated to properties of canines, biological functions, other animals, etc., but not all possible 
words anymore. Once we specify the second word ``howls''$\in \Omega_L$, context, intelligibility, and grammatical structure further 
restrict sample-space for the third word to $\Omega_M \subset \Omega_L$, from which we finally draw ``night''. Obviously, the 
nestedness in the formation of sentences is similar to the example of  the nested dice before. Nesting is imposed through 
grammatical and/or contextual, and/or interpretative constraints. 

The role of grammar for nesting is obvious. Typically in English the first word is a noun with the grammatical role of the {\em subject}.
The fact that the first word is a noun restricts the possibilities for the next word to the subset of {\em verbal phrases}.
Depending on the particular verb chosen, the words that can now follow are typically playing the grammatical role of the {\em object}, and 
are again more restricted. We use the terms sample-space reduction and nested hierarchical structure in sentences 
interchangeably. 
It is not only grammatical structure that imposes consecutive restrictions on sample-space of words as the sentence progresses, 
the need for intelligibility has the same effect. 
Without (at least partial) hierarchical structures in the formation of sentences, their {\em interpretation} 
would become very hard \cite{Partee:1976}. 
However, nested structures in sentences will generally not be strictly realised. 
Otherwise the creative use and flexibility of language would be seriously constrained. Sometimes words can act as a linguistic hinge, 
meaning that it allows for many more consecutive words, than were available for its preceding word. One expects that nestedness 
will be realised only to some degree. 
Imperfect nestedness allows for a degree of ambiguity in the linguistic code, and is one of the sources of its astonishing 
versatility \cite{Fortuny:2013}.

In this paper we quantify the degree of nestedness of a text from its word-transition matrix $M$ (network). 
To characterize the hierarchical structure of a text with a single number, we define its nestedness $n$ as a property of $M$ by 
\begin{equation}
n(M) = \left \langle \frac{|\Omega_i \cap \Omega_j|}{\min (|\Omega_i| , |\Omega_j| ) } \right \rangle_{(i,j)}  \quad ,
\label{nest}
\end{equation}
where the average is taken over all possible word pairs $(i,j)$. Nestedness is a number between 0 and 1, and specifies to what extent 
sample-space reduction is present on average in the text\footnote{  Note that the nesting indicator in Eq. (\ref{nest}) is reasonable only for the case where the probability of  two words $i,j$ having the same sample space is very low,  $p(\Omega_i=\Omega_j)\approx 0$. That is the case for the considered transition matrices.}. 
A strictly nested system, like the one shown in 
Eq. (\ref{eq:Nested}), has $n(M)=1$.
  In linguistic terms strict nestedness is clearly unrealistic.  

We use word-transition matrices from actual English texts, which serve as the input to a simple model for sentence formation. 
We then study the word frequency distributions of these artificially produced texts, and compare 
them with the distributions of the original texts. For the first time we show that it is possible to relate the topological feature of (local) nestedness in 
sentence formation to the global features of word frequency distributions of long texts. 
In this respect we propose a way to understand the statistics of word frequencies --  Zipfs law in particular -- by the actual 
structural feature of language, nestedness, without the need to resort to previous attempts including multiplicative processes, 
preferential attachment, or self-organized criticality, which,  in the context of language, sometimes seem to rest on strong and 
implausible assumptions.

\section{Model}

We assume a finite vocabulary of $N$ words. From any given text we obtain an empirical word-transition matrix $M$. Words are labeled with latin 
indices. $M_{ij}=1$ means that in the text we find at least one occasion where word $j$ directly follows $i$, if $M_{ij}=0$ word  $j$ never 
follows $i$ in the entire text. Figure 
3(a) shows the transition matrix for {\em The origin of species}. 
To quantify sample-space for individual words, note that a line $i$ in $M$ contains the {\em set} of words, $\Omega_i=\{k | M_{ik}=1\}$, that directly follow word $i$. 
By $|\Omega_i|$ we denote the size (number of elements) of $\Omega_i$, which is the number of {\em different} words that can follow $i$. $\Omega_i$ is an  
approximation for the sample-space volume that is accessible after word $i$ has occurred. 
Different words have different sample-space volumes, see Fig. 
3(b), where the sample-space profile is shown.
We parametrize the profile as $y^\kappa = x$, where $x$ corresponds to the
sample space volume, $|\Omega_i|$, and $y$ to the sample space index $i$.
We call a system {\em linearly  
nested} if $\kappa=1$  (as in Eq. (\ref{eq:Nested})), {\em weakly nested}  for $\kappa<1$ (as in Fig. 
3(b)), and {\em strongly nested} if $\kappa>1$. 
An example for a weakly nested profile can be seen in one of the insets of Fig. 4 (c).
The parameter $\kappa$ has an intuitive interpretation in terms of a measure of 'structuredness' of word-transitions. 
In the case of a weakly nested profile ($\kappa<1$) there are many words that can be followed by many different words,  
whereas in a strongly nested profile ($\kappa>1$) there are a few words that are followed by many other words, and    
many words that can only be followed by a very few. In this sense $\kappa$  measures to what extent 
word-transitions are effectively constrained.    

 Note that the profile in Fig. 
3(b) is actually not well fitted with a power law, the reason for the parametrisation is for a purely theoretical argument that will become clear below.
We exclude words that are followed by less than 2 different words in the entire text, i.e. we remove all lines $i$ from $M$ for which $|\Omega_i| < 2$. 
Strict nestedness is not to be confused with strong or weak nesting. The latter are properties of the sample-space profile.

For statistical testing we construct two randomised versions of $M$, and denote them by $M_{\rm rand}$ and $M_{\rm row-perm}$, respectively. 
$M_{\rm rand}$ is obtained by randomly permuting the rows of the individual lines of the matrix $M$. 
This keeps the number of non-zero entries in every line the same as in  the original matrix $M$,  
but destroys its nestedness and the information which words follow each other.  
The second randomized version $M_{\rm row-perm}$ is obtained by permuting the (entire) rows of the matrix $M$. 
This keeps the nestedness of the matrix  unchanged, but destroys the information on word-transitions.
\begin{figure}
\includegraphics[width= 9cm]{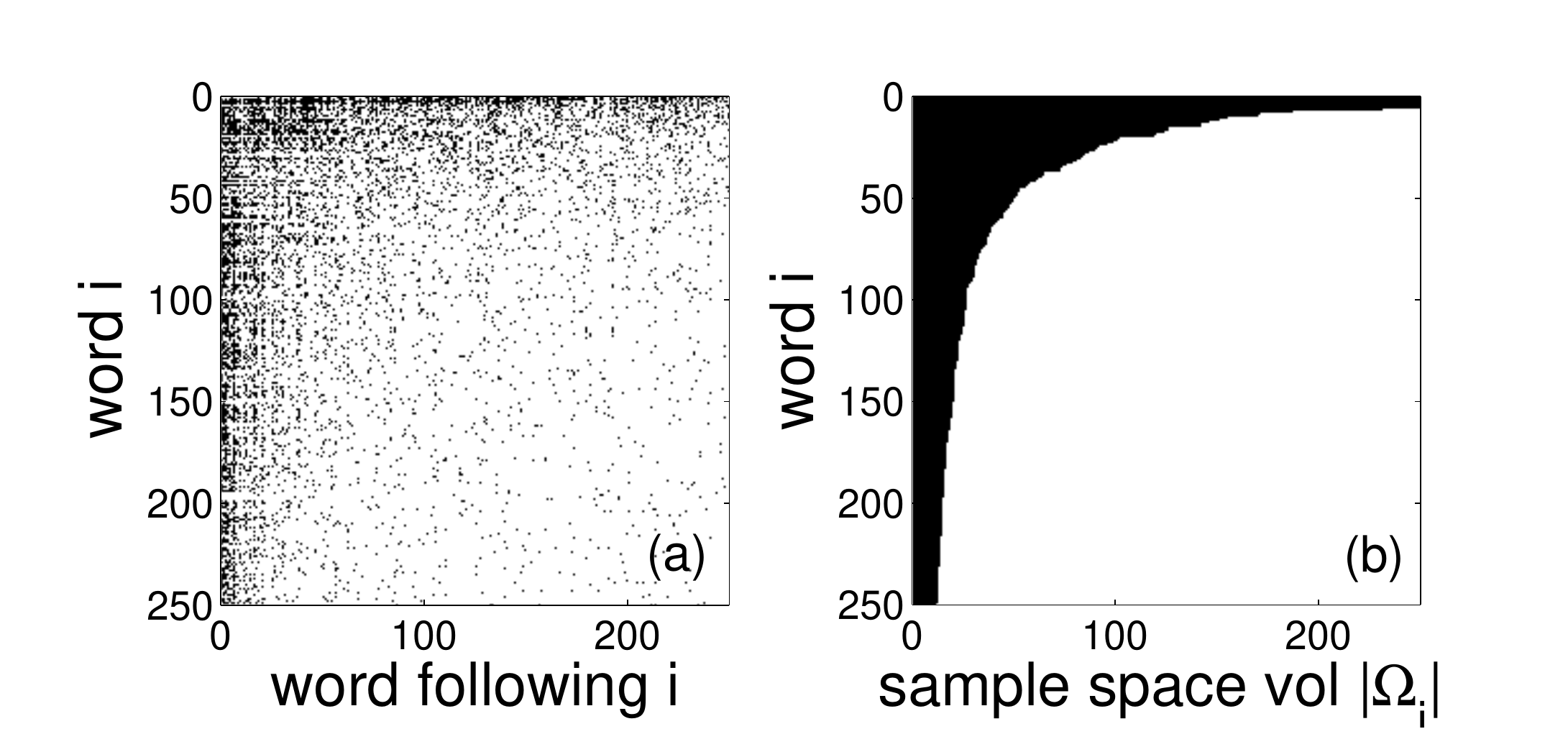}
\caption{Section of word-transition matrix $M$ for the 250 words that show the largest sample-space volume of 
consecutive words (a).  A black entry ($M_{ij}=1$) means that a given word $i$ (y axis) is followed by word $j$ (x axis). 
Non-trivial nestedness is seen by the approximate funnel like shape of the density of words. 
The actual value of the sample-space volume for every word $i$, $|\Omega_i|$ is shown in (b), 
which is obtained by shifting all entries of the lines $i$ to the leftmost positions. 
We call (b) the sample-space profile. 
}
\label{fig2}
\end{figure}
\begin{figure}
\includegraphics[width= 7.5cm]{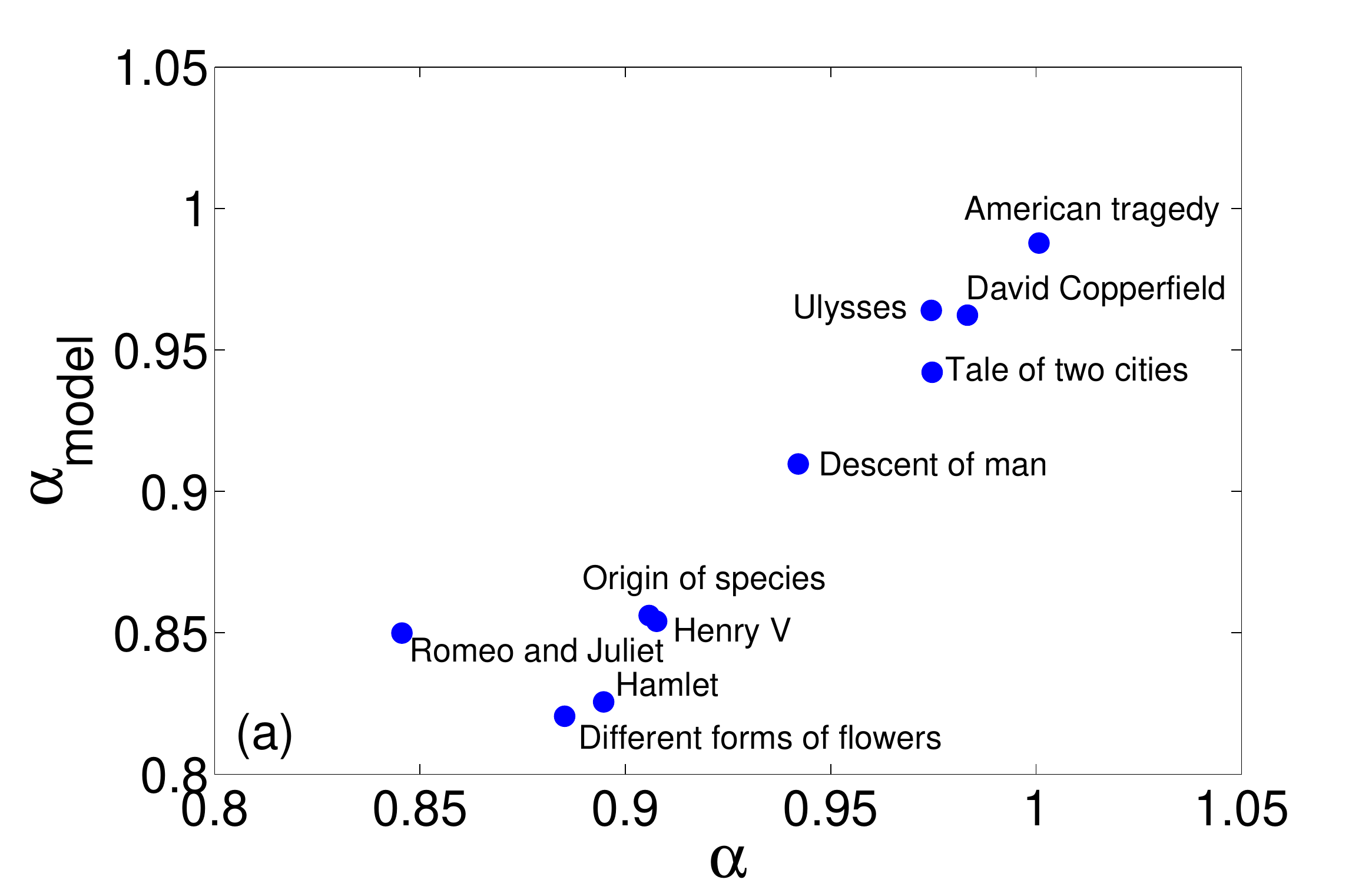}\\
\includegraphics[width= 7.5cm]{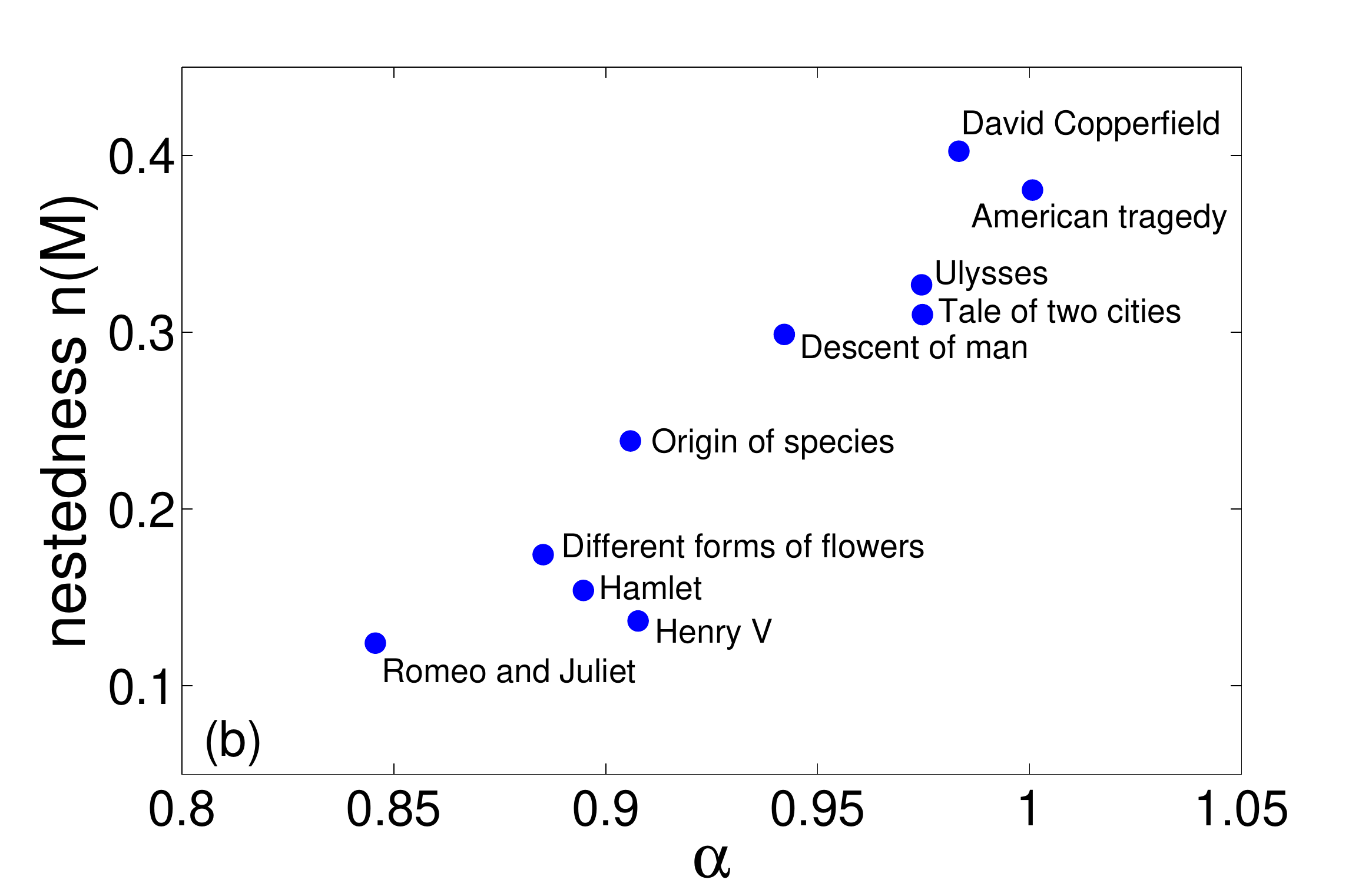}\\
\includegraphics[width= 7.5cm]{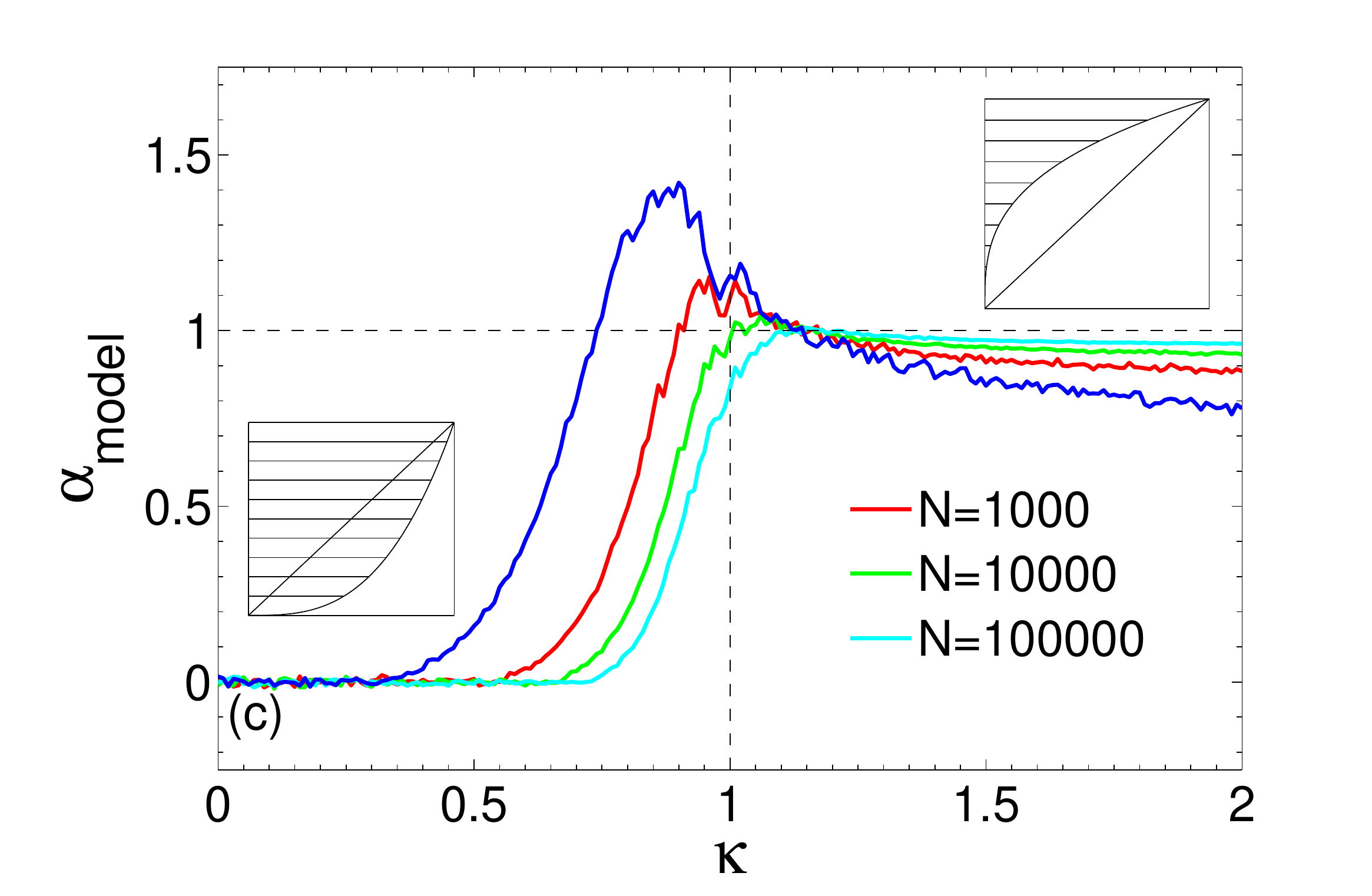}
\caption{$\alpha$ exponents from rank ordered word frequency distributions of 10 books versus model results for $\alpha_{\rm model}$ (a). 
Clearly the model explains the actual values to a large extent.
(b)  $\alpha$ exponents versus nestedness $n(M)$ of the 10 books. 
(c) $\alpha_{\rm model}$ exponents versus the sample-space profile parameter $\kappa$. For large vocabularies $N=100,000$, at $\kappa \sim 1$ a fast transition 
from the weak nesting to the strong nesting regime occurs,  where we find $\alpha_{\rm model}\sim0$ and $\alpha_{\rm model}\sim1$, respectively. 
Weak and strong nesting profiles are schematically indicated.
For smaller (realistic) $N$ the transition appears at $\kappa<1$, and $\alpha_{\rm model}$ covers a range between $\sim 0.85$ and $1.1$ in the scaling phase, which 
fits the empirical range seen in (a). 
 }
\label{fig4}
\end{figure}

Given $M$ we construct random sentences of length $L$ with the following model: 
\begin{itemize}
\item pick one of the $N$ words randomly. Say the word was $i$. Write $i$ in a wordlist $W$, so that $W=\{i\}$. 
\item jump to line $i$ in $M$ and randomly pick a word from the set $\Omega_i$. Say the word chosen is $k$; update the wordlist $W=\{i,k\}$.
\item jump to line $k$, and pick one of the words from $\Omega_k$; say you get $j$, and update $W=\{i,k,j\}$.
\item repeat the procedure $L$ times. At this stage a random sentence is formed. 
\item repeat the process to produce $N_{\rm sent}$ sentences. 
\end{itemize} 
In this way we get a wordlist with $L \times N_{\rm sent}$ entries, which is a random book that is generated 
with the word-transition matrix of an actual book.    
From the wordlist we obtain the word frequency distribution $f_{\rm model}$. 
The present model is similar to the one in \cite{Zipf1C-MHT} but differs in three aspects:  
it allows for non-perfect nesting $n<1$, it has no explicit noise component, and it has a fixed sequence (sentence) length.

\section{Results}

We analyze the model with computer simulations, specifying $L=10$, and $N_{\rm sent}=100,000$. 
We use 10 randomly chosen books\footnote{In particular we use {\em An American tragedy}, by Theodore Dreiser,  {\em The Origin of Species},  {\em Descent of Man}, and 
{\em Different Forms of Plants}, by Ch. Darwin, {\em Tale of two Cities}, and {\em David Copperfield}, by Ch. Dickens, 
{\em Romeo and Juliet}, {\em Henry V}, and {\em Hamlet}, by W. Shakespeare, and {\em Ulysses} by J. Joyce. Vocabulary varies from $N=3,102$ 
(Romeo and Juliet) to $22,000$ (Ulysses) words.} 
from the Project Gutenberg  \cite{gutenberg}. 
For every book we determine its vocabulary $N$, its matrix $M$, its $\Omega_i$ for all words, its nestedness $n(M)$, 
and the exponent of the rank ordered word frequency distribution $\alpha$
 (least square fits to $f(r)$, fit range between the $5\leq r \leq 200$). 
 $f(r)$ is seen for in Fig. 
1 (blue) the exponent is $\alpha \sim 0.90$. 
We run the model for the parameters of every individual book to generate a random text. Using the empirical $\Omega_i$ for the model ensures that 
this random text has exactly the same sample-space profile and the nestedness as the book. 

The distribution obtained from the model $f_{\rm model}$ is clearly able to reproduce the approximate power law exponent for 
{\em The origin of species}, $\alpha_{\rm model}\sim 0.86$ (same fit range). Moreover it captures details of the distribution $f$. 
For large values of $r$ in $f_{\rm model}(r)$ a plateau is forming before the exponential finite size cutoff is observed. 
Both, plateau and cutoff can be fully understood with the randomised model.

In Fig. 
4(a) we compare the $\alpha$ exponents as extracted from the books with the model results $\alpha_{\rm model}$. 
The model obviously explains the actual values to a large extent, slightly underestimating the actual exponents. 
We get a correlation coefficient of  $\rho=0.95$ ($p< 3.7\times 10^{-5}$). 
In Fig. 
4(b) we show that nesting $n(M)$ is related to the exponents $\alpha$ in an approximately linear way.    
We test the hypothesis that by destroying nestedness the exponents will vanish. Using the randomised $M_{\rm rand}$
we find $\bar \alpha_{\rm model}^{\rm rand}= 0.19 \pm 0.03$ (same fit range), which  effectively destroys the power law. 
Using the other randomized version that keeps the nestedness intact, $M_{\rm row-perm}$,   for low-rank words 
(up to approximately rank $\sim 10$)
we find similar word frequency distributions as for $M$, however, as expected, the power law tail (high ranks) vanishes for $M_{\rm row-perm}$
due to the noise contribution of the randomization (not shown).
To validate our assumption that word ordering is essential, we computed the model rank distributions by using the transposed 
matrix $M^T$, meaning that we reverse the time flow in the model. We find two results. 
First, the correlation between the exponents of the books $\alpha$, and the model $\alpha_{\rm model}^{\rm rev. time}$ vanishes, 
reflected by an insignificant correlation coefficient $\rho=0.47$ ($p=0.17$). 
Second, the exponents (averaged over the 10 books) are  significantly smaller $\bar \alpha_{\rm model}^{\rm rev. time}= 0.85 \pm 0.03$, 
than for the correct time flow, where we get $\bar \alpha_{\rm model}= 0.90 \pm 0.06$. The corresponding $p$-value of a t-test
is $0.039$.

Finally we try to understand the importance of the sample-space profile on the scaling exponents. 
For this we generate a series of $M$ matrices that have a profile parametrized with a power $\kappa$. 
In Fig. 
4(c)  the model exponents $\alpha_{\rm model}$ from these artificially generated $M$ are shown as a function of  $\kappa$, 
for various sizes of vocabulary $N$. For  $\kappa<1$ (weak nesting) we find exponents $\alpha_{\rm model}\approx0$, i.e. 
no scaling law. For large $N$ at 
$\kappa=1$ a fast transition to $\alpha_{\rm model}\approx1$ (Zipf) occurs. For smaller $N$ we find a more complicated behaviour of the 
transition, building a maximum exponent at a $\kappa<1$. The range of book exponents 
$\alpha$ ranges between $0.85$ and $1.1$, which is exactly the observed range for realistic vocabulary sizes $N\sim 1,000$-$10,000$. 
We verified that variations in sentence length (with the exception of $L=1$) do not change the reported results. 
For one-word sentences ($L=1$) we obviously get a uniform word frequency distribution, and as a consequence, a flat 
rank distribution, since most words have almost the same rank. We varied the number of sentences 
from $N_{\rm sent}=10^4$ to $10^6$, and find practically no influence on the reported results.

\section{Discussion}

In this paper we focus on the fundamental property of nestedness in any code that conveys meaningful information, such as language. 
We argue that if nesting was not present one would easily end up in confusing situations as described in {\em La Biblioteca de Babel} by J. L. Borges, 
where a hypothetical library owns all books composed of {\em all} possible combinations of characters filling 410 pages. 
We define and quantify a degree of nestedness in the linguistic code. 
Low degrees of nestedness typically imply a less strict hierarchy on word usage or a more {\em egalitarian} use of the vocabulary, than texts with 
high nestedness. 
As expected, texts have a well defined, but not strictly nested structure, which might arise from a compromise of specificity 
(to convey unambiguous messages) and flexibility (to allow a creative use of language). 
We find that nestedness  varies between different texts, suggesting that different ways of using the vocabulary and grammar are at work.
Our sample of texts included three plays by Shakespeare, three scientific texts, and four novels. We find that the plays, maybe closest to spoken language, show a lower nestedness than the science books. The novels show the highest levels of nestedness. The sample is too small to draw conclusions on whether different types of texts are characterized by typical values of nestedness, however it is remarkable that nestedness is correlated with the variations of the scaling exponents of word frequencies on a book-by-book basis.     

The main finding of this paper is that a simple sample-space reducing model can show that nestedness indeed explains the emergence of scaling laws in word frequencies, in particular, Zipf's law. 
More precisely, we were able to relate the emergence of scaling laws with topological structure of the 
word transition matrix, or ``phasespace''. The result is remarkable since the matrix   
does not encode any information about how often word $j$ follows word $i$, it just tells that $j$ followed $i$ at least once in the entire text. 
Random permutations of the matrix that destroy its nestedness can not explain the scaling anymore, while permutations that keep nesting intact, do indicate the existence of the power laws.  
It is further remarkable that no (non-local) preferential, multiplicative, or self-organized critical assumptions are needed to understand the observed scaling, 
 and that no parameters are needed beyond the word transition matrices. 
 
The fact that the simple model is so successful in reproducing the detailed scaling property in word frequency statistics 
might point to an important aspect of language, that has not been noted so-far; the fact that overall 
word-use is statistically strongly influenced by the use of local hierarchical structures and constraints that we use in generating sentences. 
We believe that the close relation between nestedness and the scaling exponent opens the door for an interpretation of word frequency distributions 
as a statistical observable that strongly depends on the usage of the vocabulary and grammar within a language. 
Accordingly we conjecture that Zipf's law might not be universal, but that word-use statistics depends on local structures which may be different across texts and 
even within sentences. Further research is needed to clarify this point.

Finally, it is worthwhile to note that the class of sample-space reducing processes provide an independent route to scaling 
that might have a wide range of applications for history-dependent and ageing processes \cite{Zipf1C-MHT}.   
In statistical physics it is known that processes that successively reduce their phasespace as they unfold  
are characterised by power law or stretched exponential distribution functions. These distributions generically 
arise as a consequence of phasespace collapse \cite{hanel2014}.

{\bf Acknowledgements}$\\$
This work was supported by the Austrian Science Fund FWF under KPP23378FW.

{\bf Competing financial interests}$\\$
The authors declare no competing financial interests.

{\bf Author Contributions}$\\$
ST designed the research, performed numerical analysis and wrote the manuscript. RH and BC-M performed numerical analysis and wrote the manuscript.
BL did preprocessing the books, and performed numerical analysis.

\end{document}